\documentclass{elsart}
\usepackage[dvips]{graphics}
\usepackage{amsmath}
\def\units#1{\hbox{$\,{\rm #1}$}}

\begin{document}
\begin{frontmatter}

\title{A Monte Carlo code for full simulation of a transition
radiation detector}
\author{M.N. Mazziotta\thanksref{mnm}}
\address{Dipartimento di Fisica dell'Universit\'a and INFN Sezione di
Bari, via Amendola, 173, I-70126 Bari (Italy)}

\thanks[mnm]{fax: +39 080 5442470; e-mail: mazziotta@ba.infn.it}

\begin{abstract}
A full simulation of a transition radiation detector (TRD) based
on the GEANT, GARFIELD, MAGBOLTZ and HEED codes has been developed. 
This simulation can be
used to study and develop TRD for high energy particle identification
using either the cluster counting or the total charge measurement
method. In this article it will be also shown an application of this
simulation to the discrimination of electrons
from hadrons in beams of momentum of few $GeV/c$ or less, assuming typical
TRD configuration, namely radiator--detector modules.
\end{abstract} 


\end{frontmatter}
\begin{center}
{\em(To be submitted to Computer Physics Communication)}
\end{center}

\section{Introduction} 
\label{sec:intr}
Transition radiation (TR) is an electromagnetic radiation produced by
ultrarelativistic charged particles crossing the interface between two
materials with different dielectric properties \cite{Ginz,Gari}. The TR
spectrum is peaked in the X-ray region and the probability 
of a X-ray photon being emitted at each interface is of the order 
of $\alpha \simeq1/137$. 
The transition radiation yield is proportional to the Lorentz
factor $\gamma$ of the incident charged particle and is
independent on the kind of particle. 
That offers an attractive alternative to identify particles
of given momentum with a non destructive method. 

In order to enhance the TR X-ray production, radiators
consisting of several hundred foils regularly spaced or irregular
radiators of few $cm$ of thickness consisting of carbon compound foam
layers or fiber mats are usually adopted.
The ``multilayer'' radiator introduces significant
physical constraints on the radiation yield, because of the so-called
``interference effects''. It has been established that the radiation
emission threshold occurs at a Lorentz factor $\gamma_{th}=2.5 \omega_p
d_1$, where $\omega_p$ is the plasma frequency (in \units{eV} units) of
the foil material, and $d_1$ is its thickness in \units{\mu m}. For
$\gamma \geq \gamma_{th}$ the radiation yield increases up to a
saturation value given by
$\gamma_{sat} \sim \gamma_{th} (d_2/d_1)^{1/2}$, where $d_2$ is the width
of the gap between the foils \cite{Artr}. 

The conventional method of TR detection is the measurement of the sum of
the energy released by ionization and from photoelectrons produced by
TR X-rays. 
The radiating particle, if not deflected by magnetic fields, releases its
ionization energy in the same  
region as the X-ray photons, introducing a background signal 
that can be reduced if a gaseous detector is used. 
Since the gas must provide efficient conversion of the TR photons,
the use of high-Z gases is preferred.
The detector usually consists of proportional chambers filled with argon
or xenon with a small addition of quenching gases for gain 
stabilization ($CO_2$, $CH_4$).

The measurement of TR using proportional chambers is   
generally based on one or both of the following methods:
\begin{itemize}
\item the ``charge measurement'' method, where the signal collected from
a chamber wire is charge analyzed by ADCs \cite{Fisc};
\item the ``cluster counting'' method,
where the wire signal is sharply differentiated in order to discriminate
the X-ray
photoelectron clusters producing pulses (hits) exceeding a threshold amplitude
from the $\delta$-ray ionization background \cite{Ludl}.
\end{itemize}
In both cases a cut on the analyzed charge or on the number of
clusters is needed in order to
discriminate radiating particles from slower nonradiating
ones. Multiple module TRDs,
with optimized gas layer thickness, are normally employed to
improve background rejection. A reduced chamber gap limits the particle
ionizing energy losses, while the
X-rays escaping detection may be converted in the downstream chambers.

Transition radiation detectors are presently of interest in fast particle
identification, both in accelerator experiments \cite{Camp,Dolg} 
and in cosmic ray physics \cite{Prin}-\cite{Barb2}. A TRD is used
to evaluate the underground cosmic ray muon energy spectrum
in the Gran Sasso National Laboratory \cite{TRDMAC}. 
In spite of their use in several high energy
experiments, a simulation code is not yet available in the
standard simulation tools. 

Several codes based on parameterizations  of test beam measurements
have been developed to simulate the TRDs
\cite{Cast,atlas}. Lately a TRD has been proposed in a Long 
Base Neutrino Oscillation Experiment \cite{noe}, in which a simulation
has been developed using a GEANT interface \cite{gnoe}. The results achieved
in the last experience
have been rather satisfactory, in spite of some difficulties to track low
energy photons in GEANT.
 
In this paper a full simulation of a TRD is described. The program is
based on GEANT \cite{Brun}, GARFIELD \cite{Garf}, 
MAGBOLTZ \cite{Magb} and HEED \cite{Heed} codes 
in order to exploit the best performances in each one. 
In this way a full simulation has been developed tracking the particles into
the detector and producing the pulse shape from each proportional tubes. 

\section{Transition radiation emission}
\label{TRE}
Extensive theoretical studies have been made about TR. 
The basic properties of the TR production as well as the interference 
phenomena in multifoil radiator stacks are rather well understood
and well described with classical electromagnetism
(for instance  see \cite{Fabj1}). There was also an
attempt to give a quantum description of TR \cite{Gari1}.
The quantum corrections to the TR intensity become interesting
for the emission of very high energy photons, namely when the TR
photon energy is comparable with the energy of the radiating
particle. Therefore they are no longer
significant in the X-ray region for incident charged  particle of momenta
of few $GeV/c$ and the  
expressions derived are similar to the classical theory. Therefore, the TR 
emission is described for practical purposes by classical formulation,
and the TR energy is considered carried out by photons (quanta).

As shown by Artru et al. \cite{Artr} the TR energy $W$ emitted from a
stack of $N$ foils of thickness $d_1$ at regular distances $d_2$,
without taking into account absorption effects,
can be written as:
\begin{equation}
\cfrac{d^2 W}{d \omega~d \theta^2} =  
\eta~4~\sin^2{\cfrac{\phi_1}{2}} 
\left(\dfrac{\sin{N~\cfrac{\phi}{2}}}{\sin{\cfrac{\phi}{2}}} \right)^2
\label{e1}
\end{equation}
Where
\begin{equation}
\eta = \frac{\alpha}{\pi} \left( 
\frac{1}{\gamma^{-2}+\theta^2+\xi_1^2} -
\frac{1}{\gamma^{-2}+\theta^2+\xi_2^2} \right)^2 \theta^2
\label{e2}
\end{equation}
is the energy emitted at each interface. In eq. (\ref{e1}) and
(\ref{e2}) $\theta$ is the angle between
the incident particle and the TR X-ray, and $\xi_i=\omega_i/\omega$ where
$\omega$ is the TR quantum energy (in \units{eV} units)
and $\omega_i$ are the plasma energies
of the two media ``1'' (foil) and ``2'' (gap).

The factor $4~\sin^2{\cfrac{\phi_1}{2}}$ in eq. (\ref{e1}) 
is due to the coherent superposition
of TR fields generated at the two interfaces of one foil,
with the phase angle
$\phi_1=d_1/z_1$ being the ratio of the foil thickness $d_1$ 
(in \units{\mu m} units)
to the ``formation zone'' $z_1$ of the foil material:
\begin{equation}
z_1 = \left( 2.5~\omega~(\gamma^{-2}+\theta^2+\xi_1^2) \right)^{-1}
\label{e3}
\end{equation}
The last factor of eq. (\ref{e1}) describes the coherent interference 
of TR in a stack composed
of $N$ foils and gaps at regular distances $d_2$. $\phi=\phi_1+\phi_2$ 
is the total phase angle of one basic foil plus gap, with $\phi_2$ being 
defined in analogy to $\phi_1$. The TR X-ray energy distribution can
be obtained by taking the ratio of equation (\ref{e1}) to $\omega$.

Since the TR yield from multifoil stack is described as an
interference phenomenon due to whole radiator, in order
to calculate the total TR quanta emitted by the particle
crossing the radiator, one needs to known the total number of foils crossed.
Therefore it is not possible to follow the particle
into radiator in order to calculate the probability to emit a quantum
in a given step, i.e. we do not have a cross section for the TR effect. 
That may introduce some difficulties to simulate the TR process.
Moreover, the TR intensity is a complex function
of the thicknesses $d_1$ and $d_2$, of the plasma energies $\omega_1$ and 
$\omega_2$ for a given $\gamma$ Lorentz factor. This behaviour may introduce
an additional difficulty to calculate the TR spectra for any kind of
radiators.

The energy of the TR photons depends on the radiator material and its
structure. In ref. \cite{Artr} it is shown that the average TR energy
carried
out by quanta is given by:
\begin{equation}
< \omega > \simeq 0.3~\gamma_{th}~\omega_1
\label{e4}
\end{equation}
Assuming $d_1=10~\mu m$ and $\omega_1=20~eV$ one obtains
$\gamma_{th} \sim 500$ and $<\omega> \sim 3~keV$.
This may introduce some difficulties to track soft X-ray photons in a
medium.

The ability to identify particles by a TRD is determined by the relative
amounts of TR and ionization energy loss in the proportional chambers.
Large fluctuations of ionization loss in thin gas layers limit this
methods. Therefore, in order to better understand the performance of a TRD,
one needs careful calculations of ionization energy loss and its fluctuations,
producing knock-on or $\delta$-electrons. On the other hand,
if one would like to use the
cluster counting method to separate the TR X-ray from the track ionization
background, then the range and the size of $\delta$-electron and of
photoelectron, the number of electron--ion pairs 
produced in the gas and their arrival
time on the wire need to be taken into account. Finally the current produced
on the anode wire of the gas chambers and its pulse shape fed to discriminator
by the front end electronics also play an important role in this method.

\section{TRD full simulation}
On the basis of the above discussion, the approach followed to simulate a
TRD is based on the codes GEANT, GARFIELD, MAGBOLTZ and HEED (the last
two codes are used by GARFIELD). The geometric description
of the detector has been given by GEANT, including the simulation of all
physical processes that occur in the materials crossed by the particles. The 
ionization energy loss and the photoelectric process in the gas have been 
not considerated in the GEANT code, because they are simulated by HEED. 

When charged particles cross the gas of proportional chambers,
or photons are entering into these volumes, the HEED package is called.
In this way the ionization energy loss and the electron--ion pairs distribution
along the track are calculated. The photoelectric absorption of photons
in the gas is also simulated, including the evaluation of the
photoelectrons produced and the total number of electron--ion pairs. 
Finally the current pulse produced on the anode
wire is evaluated by the GARFIELD code using the gas properties as
its drift velocity and gain calculated by
the MAGBOLTZ program as a function of the electric field.

\subsection{TR process}
The GEANT code does not simulate transition radiation.
In order to produce the TR photons in GEANT, a physical
process has been introduced whenever a relativistic charged particle
crosses the radiator. 

The TR photon energy spectrum and the mean number
of X-ray are calculated for the input radiator and for the 
energy of primary particle which one simulates.
When the charged particle crosses the radiator and comes out the TR
process 
is activated. The total number of TR photons is generated according to 
a Poisson distribution if their average number is less than 10, otherwise 
a Gaussian distribution may be used. The energy of each TR X-ray is randomly
generated according to a calculated spectrum and its position is
generated along the radiating
particle path at the end of a radiator. The produced TR photons are
then treated as secondary particles in GEANT and they are stored in the
common block GCKING. 
In order to be transported by GEANT, these photons are stored
in the data structure JSTAK by the GSKING routine.

\subsubsection{TR formulas used in the code}
The TR production relations used in this simulation take into account
the photon absorption in the radiator. This effect has been simulated
using the GEANT absorption lengths of the photons calculated for this
material.
\\

\noindent {\bf Regular radiator} \\
The energy distribution of TR photons for a stack of plates 
taking into account the absorption in the foils and gaps is given 
by \cite{Artr}:
\begin{equation}
\cfrac{d^2 N}{d \omega~d \theta^2} = \cfrac{1}{\omega} 
~\eta~4~\sin^2{\cfrac{\phi_1}{2}} 
\left(\cfrac{\sin^2{N~\cfrac{\phi}{2}}+\sinh^2{N~\cfrac{\sigma}{2}}}
{\sin^2{\cfrac{\phi}{2}}+\sinh^2{\cfrac{\sigma}{2}}}  \right)
\e^{-\frac{N-1}{2}~\sigma}
\label{e5}
\end{equation}
where $\sigma=d_1 / \lambda_1 + d_2 / \lambda_2$ is the absorption in one
foil + one gap and $\lambda_1$ and $\lambda_2$ are the absorption
lengths for the emitted radiation in two media as calculated by GEANT
(see paragraph \ref{p3.2}).
 
For large values of the number of foils $N$, the $\delta$ function can
be assumed to approximate the last two factors of the above expression.
Making this approximation and integrating over $\theta^2$, equation 
(\ref{e5}) becomes:
\begin{equation}
\cfrac{d N}{d \omega} = \cfrac{1}{\omega}~ \cfrac{4~\alpha~N_{equ}}{1+\tau}
\sum_n \theta_n \left( \cfrac{1}{\rho_1+\theta_n}-\cfrac{1}{\rho_2+\theta_n}
\right)^2 \left(1-\cos{(\rho_1+\theta_n)}\right)
\label{e6}
\end{equation}
where \\ \\
$\rho_i=2.5~d_1~\omega~(\gamma^{-2}+\xi_i^2)$; \\ \\
$ \tau=d_2/d_1$; \\ \\
$\theta_n=\cfrac{2~\pi~n-(\rho_1+\tau \rho_2)}{1+\tau}>0$; \\ \\
$N_{equ}=\cfrac{1-\e^{-N \sigma}}{1-\e^{-\sigma}}$~~. \\ \\
$N_{equ}$ is the number of
equivalent foils when the absorption is take into account.

\begin{figure}[h]
\resizebox{14.0cm}{14.0cm}{\includegraphics{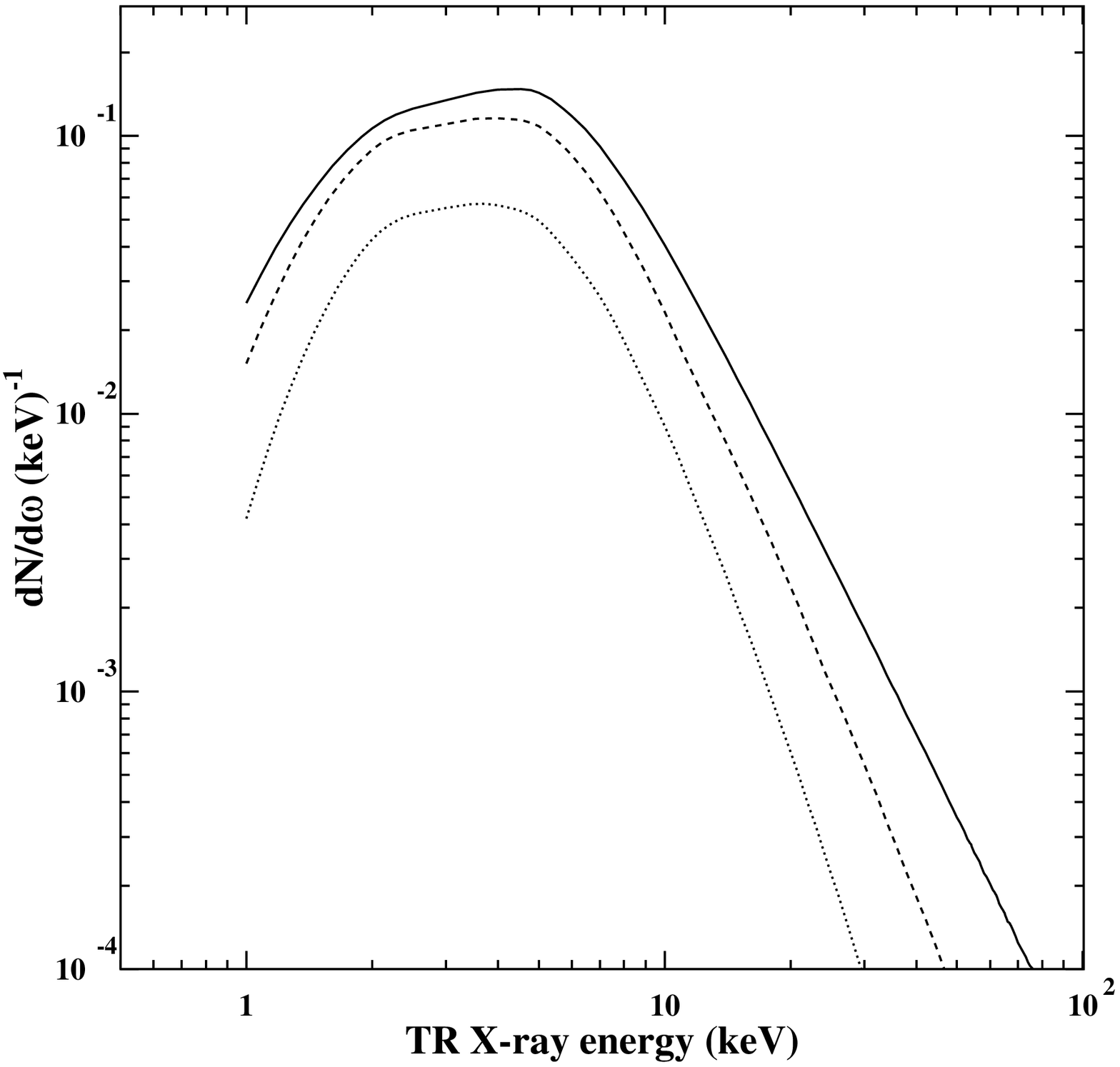}}
\caption{The TR spectra generated by 250 foils of polyethylene
($d_1=5~\mu m$ and $\omega_1=20~eV$) at regular distances $d_2=200~\mu
m$ in air ($\omega_2=0.7~eV$). Solid line: $\gamma=5000$; dashed line:
$\gamma=1000$ and dotted line:  $\gamma=500$. }
\label{trreg}
\end{figure}

To evaluate the total number of TR photons the numerical calculation
of equation (\ref{e6}) has been carried out at selected X-ray energies 
($\omega$), from
$1~keV$ to $100~keV$, with a precision better than $10^{-3}$.
%\begin{figure}[h]
%\resizebox{14.0cm}{9.0cm}{\includegraphics{trreg.eps}}
%\caption{The TR spectra generated by 250 foils of polyethylene
%($d_1=5~\mu m$ and $\omega_1=20~eV$) at regular distances $d_2=200~\mu
%m$ in air ($\omega_2=0.7~eV$). Solid line: $\gamma=5000$; dashed line:
%$\gamma=1000$ and dotted line:  $\gamma=500$. }
%\label{trreg}
%\end{figure}
In Fig. \ref{trreg} the TR spectra for a regular radiator, evaluated
taking into account the absorption in the radiator, are shown.
They are calculated from the eq. (\ref{e6}). This figure shows
a broad peak around $3-5~keV$ energy, corresponding to TR mean energy 
produced by the regular radiator adopted.\\

\noindent {\bf Irregular radiator} \\
The transition radiation has been observed in irregular materials
consisting for instance
of plastic foams. A general formulation of the spectral distribution
of the number of TR X-ray quanta produced in a irregular medium, consisting
of randomly parralel plates of arbitrary thickness, is given by
Garibian et al. \cite{Gari2}. This formulation has been given with the plates
arranged in vacuum. It has been modified to take into account the
presence of a material in the gap.

The average number of radiation quanta taking into account the
absorption of the radiation is given by: 
\begin{equation}
<\cfrac{d^2 N}{d \omega~d \theta}> = \cfrac{2~\alpha}{\pi~\omega} 
\left( \cfrac{1}{1-\beta^2~\epsilon_1+\theta^2} -
\cfrac{1}{1-\beta^2~\epsilon_2+\theta^2} \right)^2 \theta^3 I
\label{e7}
\end{equation}
Here
\begin{eqnarray}
I=2~\cfrac{1-p^N}{1-p}~
Re\cfrac{(\cfrac{1+p}{2}-h_1)-(p-h_1~\cfrac{1+p}{2})~h_2}{1-h_1~h_2}+ 
& \\
2~Re\cfrac{(1-h_1)~(p-h_1)~h_2~(p^N-h_1^N~h_2^N)}{(1-h_1~h_2)~(p-h_1~h_2)} 
& \nonumber
\label{e8}
\end{eqnarray}
is the factor due to the superpositions of the radiation fields in the
plates and in the gap. The other parameters are: \\ \\
$\epsilon_k=1-(\omega_k/\omega)^2+i/(5~\lambda_k~\omega)$; \\ \\
$h_k=<\e^{-i~\phi_k~d_k}>$;  \\ \\
$\phi_k=5~\omega~\left( \beta^{-1}-\sqrt{\epsilon_k-\sin^2{\theta}} \right)
=\phi'_k + i \phi''_k$; \\ \\
$p=<\e^{-d_1/\lambda_1}>~<\e^{-d_2/\lambda_2}>$. \\ \\
The angle brackets denote the
averaging of random quantities with a distribution determined by the
distributions of $d_1$ and $d_2$.

For most of foam radiators the random foil and the gap thickness
can be described by a gamma distribution \cite{Fabj}.
In this way one finds that \cite{Gari2}: \\ \\ 
$h_k=|h_k|~\e^{i~\psi_k}$; \\ \\
$|h_k|=\left( 
\left( 1 +  \cfrac{<d_k>}{2~\lambda_k~\alpha_k} \right)^2 + 
\left( \cfrac{\phi'_k~<d_k>}{\alpha_k} \right)^2 \right)^{-\alpha_k/2}$; \\ \\
$\psi_k=-\alpha_k~arctg~\cfrac{\phi'_k~<d_k>} 
{\alpha_k+<d_k>/(2~\lambda_k)}$; \\ \\
$p=\left( 1 +  \cfrac{<d_1>}{\lambda_1~\alpha_1}
\right)^{-\alpha_1}~\left( 1 + \cfrac{<d_2>}{\lambda_2~\alpha_2}
\right)^{-\alpha_2}$. \\ \\
The parameters $\alpha_k$ represent the degree of irregularity:
\mbox{$\alpha_k=(<d_k>/\sigma_k)^2$} 
where $<d_k>$ and  $\sigma_k$ are the mean values and the mean squared 
deviations respectively of foil ($k=1$) and gap ($k=2$) thickness
distributions.

\subsection{Use of the GEANT package}
\label{p3.2}
The GEANT 3.21 code is used to describe the geometrical volumes inside
the detector and to define the materials. It has been done by the standard 
GEANT routine taking care of tracking parameters in order to define the
active physical processes and the cuts (GSTPAR). In this way, the photons are
tracked using the GEANT absorption coefficients and the gamma cuts have
been lowered to $1~keV$ in all the materials.

The materials used, which are not defined in the default GEANT
program, have been
implemented using the standard routine (GSMATE or GSMIXT). The radiators
have been defined as a mixture composed by the foil material and the
gap material (air) containing the proportion by weights of each material.
The foil materials and the gas chamber walls have been defined as compounds
containing the proportion by number of atoms of each kind \cite{Brun}. 

\begin{figure}[h]
\resizebox{14.0cm}{14.0cm}{\includegraphics{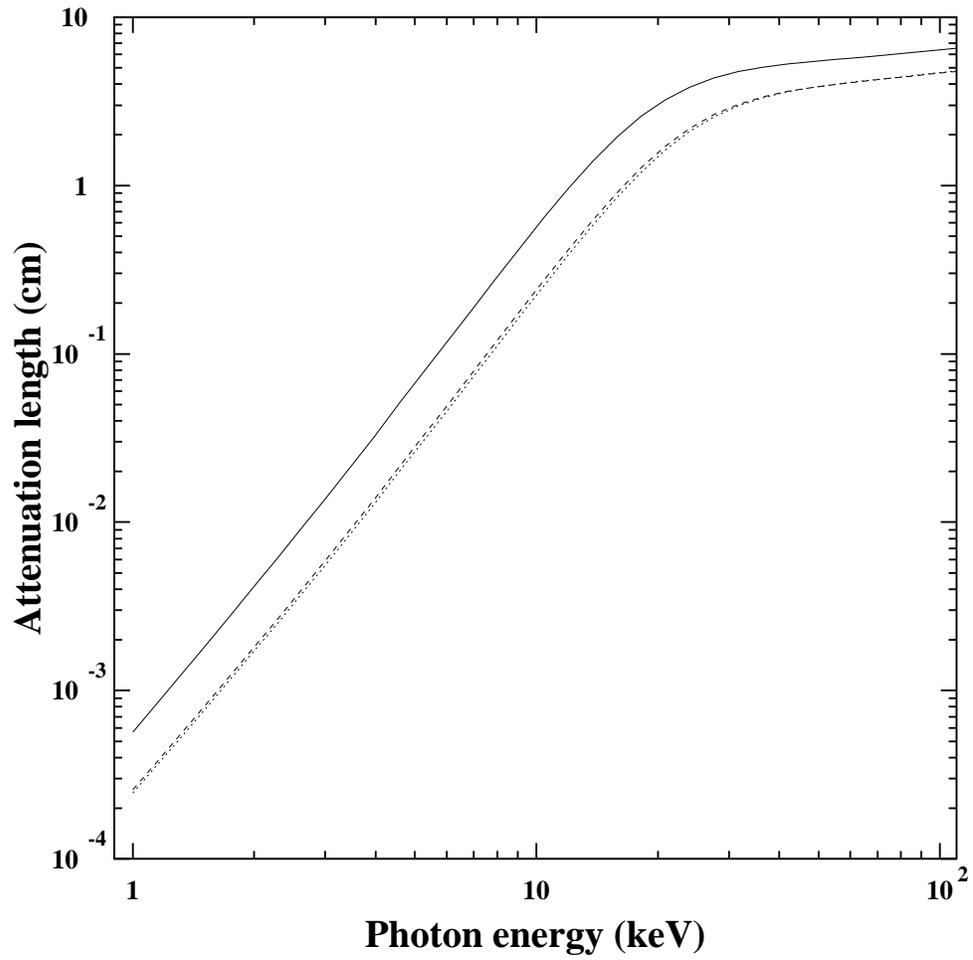}}
\caption{Photon attenuation length for different materials
as calculated by GEANT routines in the range from $1~keV$ to $100~keV$.
Solid line: polyethylene; dashed line: kapton and dotted line: mylar.}
\label{phlam}
\end{figure}

In Fig. \ref{phlam} the photon attenuation
lengths calculated by GEANT 
for polyethylene ($C H_2$, $\rho=0.93~g/cm^3$),
kapton ($C_{22} H_{10} N_2 O_5$, $\rho=1.42~g/cm^3$) 
and mylar ($C_5 H_4 O_2$, $\rho=1.05~g/cm^3$) 
are shown.
In this figure one can see that the kapton photon attenuation 
length is always less than polyethylene and the photon attenuation 
length for kapton is the same as for mylar.
%\begin{figure}[h]
%\resizebox{14.0cm}{9.0cm}{\includegraphics{totphot.eps}}
%\caption{Photon attenuation length for different materials
%as calculated by GEANT routines in the range from $1~keV$ to $100~keV$.
%Solid line: polyethylene; dashed line: kapton and dotted line: mylar.}
%\label{phlam}
%\end{figure}

The gas chambers are the sensitive volume of the TRD and for each charged
particle crossing the gas or for each photons absorbed inside, a GEANT
HITS structure is defined to describe the interaction between particle and
detector. In the HITS structure the following information are stored:
\begin{itemize}
\item HITS(1) = number of volume level (by GEANT);
\item HITS(2) = energy loss in the gas (by HEED);
\item HITS(3) = input time in the volumes (by GEANT);
\item HITS(4:6) = x, y and z of entry point in the volume (by GEANT);
\item HITS(7:9) = x, y and z of exit point in the volume (by GEANT);
\item HITS(10) = number of cluster produced in the gas (by HEED);
\item HITS(11) = number of electron--ion pairs produced in the gas (by HEED);
\item HITS(12:111) = current pulse on the wire for 100 time slices
(by GARFIELD).
\end{itemize}
The DIGIT structure is similar to the HITS one, where the information
are stored as a sum of all particles crossing that volume, while the input
and the output coordinate are relative to the primary particle which has 
crossed the chamber.

The event processing is a highly CPU consuming job. To optimize CPU usage
DST files are produced to be analyzed at a later time. For each
event the GEANT ZEBRA data structures containing the geometrical definition,
the input kinematics, the tracking banks (JXYZ) and 
the simulated detector response (HITS and DIGIT banks) are stored in DST
files which provide the input data set for the analyses to be performed.
In this way, the electronic response of the chamber front end can be
implemented starting by the anode current impulse.   
In order to save some run informations the HEADER bank is also used by the
GSRUNG routine.

\subsection{Use of the GARFIELD package}
The GARFIELD program has been developed to simulate gaseous wire
chambers operating in proportional mode.
It can be used for instance to calculate the field maps
and the signals induced by charged particles, taking both 
electron pulse and ion tail into account.
An interface to the MAGBOLTZ program is provided for the computation of 
electron transport properties in nearly arbitrary gas mixtures. 
Starting from version 6, GARFIELD has also an interface with
the HEED program to simulate ionization of gas molecules by particles 
traversing the gas chamber.
A few examples of GARFIELD results can find via WWW \cite{Garf,Heed}. 

The HEED program computes in detail the energy loss of fast charged
particles in
gases, taking $\delta$-electrons and optionally multiple scattering of the
incoming particle
into account. The program can also simulate the absorption of photons through
photo-ionization in gaseous detectors. 
From this program, the distribution of electron--ion pairs 
along the particle track
length in the gas has been computed by GARFIELD.
Some modifications have been included in
the GARFIELD default version in order to calculate 
the cluster size distribution of photons absorbed in the gas by HEED.
Starting from these cluster size distributions the current anode wire signal
is calculated by GARFIELD. 

\begin{figure}[h]
\resizebox{14.0cm}{14.0cm}{\includegraphics{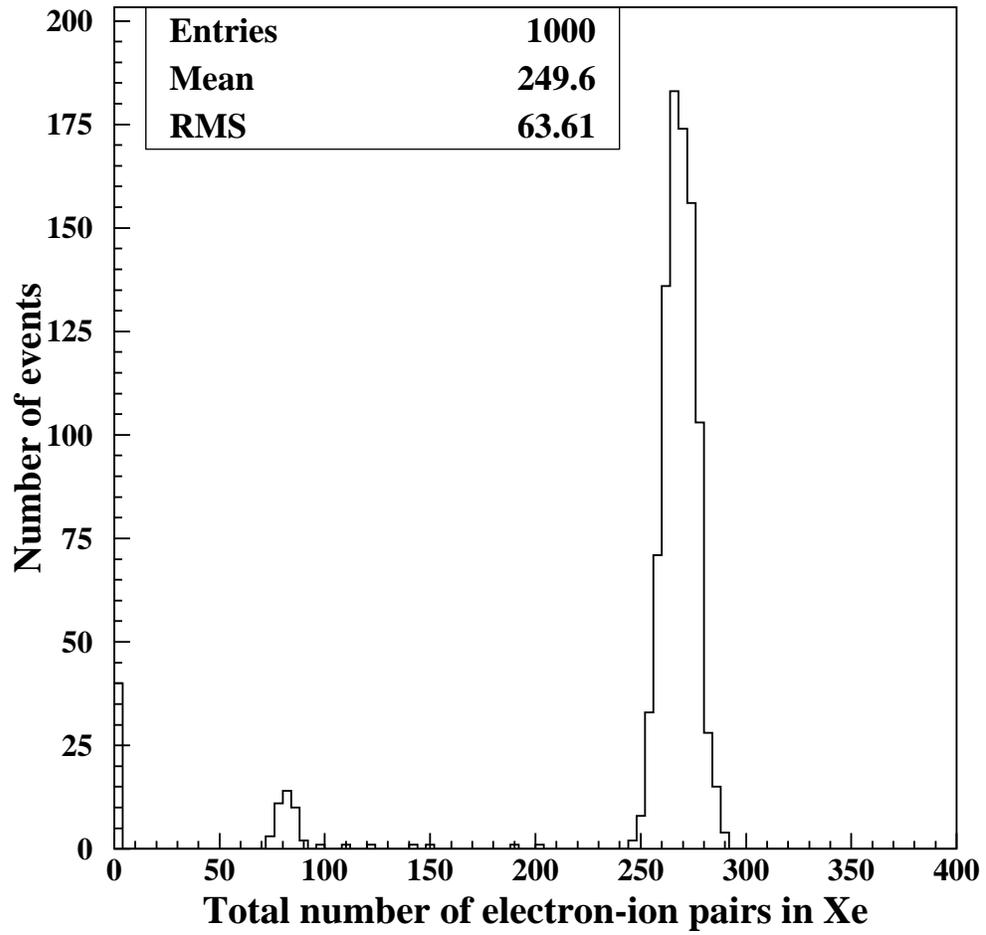}}
\caption{Electron--ion pairs distribution for $1~cm$ of xenon at NTP produced
by photons of $5.9~keV$ (\nuc{55}{Fe}).}
\label{pair}
\end{figure}

In Fig. \ref{pair} the pair distribution produced by $5.9~keV$ photons
(\nuc{55}{Fe}) in $1~cm$ of xenon at NTP is shown. From this figure one can
see the presence of a mean peak of about 270 electron--ion pairs due to 
the photoelectron
and the Auger electron. There is also a secondary peak due to 
occasional detection of a photoelectron whitout Auger emission.
%\begin{figure}[h]
%\resizebox{14.0cm}{9.0cm}{\includegraphics{fe55_xe_pair.eps}}
%\caption{Electron--ion pairs distribution for $1~cm$ of xenon at NTP produced
%by photons of $5.9~keV$ (\nuc{55}{Fe}).}
%\label{pair}
%\end{figure}

In this TRD simulation the GARFIELD 6.27 version has been used.
From the source files of GARFIELD program, written in Fortran 77 and Patchy
as pre-processor, the main routines have been included in the code
together with the GEANT routines. Some modifications have been introduced
in order to
skip interactive input information used by GARFIELD.  
All information to run the program are given via FFREAD data card.
The cell definition and the gas composition of the chambers to be
simulated have been processed in initialization of the program.

\section{Program description}
The main items of this simulation have already been 
described in the above discussion.
In this section an example of how the program works is given. 
It has been written
in Fortran by patchy as pre-processor on a PC $166~MHz$, $80~MB$ of RAM,
in the LINUX system (RedHat 5.2 version). It is transportable on any system 
changing some patchy control flags in cradle files. 

There are two codes: the first is 
dedicated to event simulation for DST production;
the second one is used to analyze the DST files including a graphical
interface too.
The input of these program is given via data cards by FFREAD
facility. The user inputs for the first program
are stored in the run header bank after the
initialization to be used by the second one.

\subsection{Geometry}
The geometry used to simulate a TRD consists of 10
radiator-proportional chamber modules. The radiator consists of 250
polyethylene foils of $5~\mu m$ of thickness at regular distances of 
$200~\mu m$ in air. The chamber consists of two planes of 16 cylindrical
proportional tubes each of $2~mm$ of radius (straw tubes) to form a
double layer close pack configuration.
These tubes are widely used in recent high
energy physics experiments \cite{atlas,Barb3}. Since the typical materials
used for the tube wall are made by carbon compounds (kapton,
mylar and polycarbonate) and their thickness are typically 
$30-50~\mu m$, the straw tubes are good candidate to be used as X-ray
detector due to the reduced attenuation length of the wall.

In this simulation the straw tube walls are made of kapton of $30~\mu
m$ thickness internally coated with copper of $0.3~\mu m$ thickness. 
The anode wire used is of $25~\mu m$ thickness. The gas mixture used is 
based on $Xe(80\%)-CO_2(20\%)$ at atmospheric pressure. The anode
voltage used is $1450~Volt$ which corresponds to a gas gain of about
$2\cdot10^4$.

\subsection{Front end electronic}
The front end electronic used in this simulation consists of a simply
amplifier which is described by a low band-pass transfer function
with a bandwidth of
$50~MHz$ and an overall gain of 10:
\begin{equation}
\tilde{A}(\omega) = A_0 \cfrac{1}{1+i \cfrac{\omega}{\omega_0}}
\end{equation}
where $A_0=10~mV/\mu A$ and $\omega_0=50~MHz$.

The anode current produced in the proportional tubes as a function of
the time $I(t)$ is converted in the output voltage amplitude $V(t)$ by:
\begin{equation}
V(t) = \int_{t}^{\infty} I(t')~A(t-t')~dt'
\end{equation}
where $A(t)$ is the Fourier transform of $\tilde{A}(\omega)$:
\begin{equation}
A(t)= \left\{ \begin{array}{ll}
              \omega_0~\e^{-\omega_0~t} &,~if~ t \geq 0 \\
              0                         &,~otherwise
             \end{array} \right.
\end{equation}
In this example no noise is assumed. 
Of course a real electronics is described by a more complex transfer
function with an electronic noise.

%\begin{figure}[h]
%\resizebox{14.0cm}{9.0cm}{\includegraphics{fe55cur.eps}}
%\caption{Anode current signal 
%produced by a X-ray of $5.9~keV$ absorbed in a tube.}
%\label{fecur}
%%\end{figure}
%%
%%\begin{figure}
%\resizebox{14.0cm}{9.0cm}{\includegraphics{fe55vol.eps}}
%\caption{Output amplitude voltage
%produced by a X-ray of $5.9~keV$ absorbed in a tube as processed by
%the low band-pass electronic.}
%\label{fevol}
%\end{figure}

\begin{figure}[h]
\resizebox{14.0cm}{14.0cm}{\includegraphics{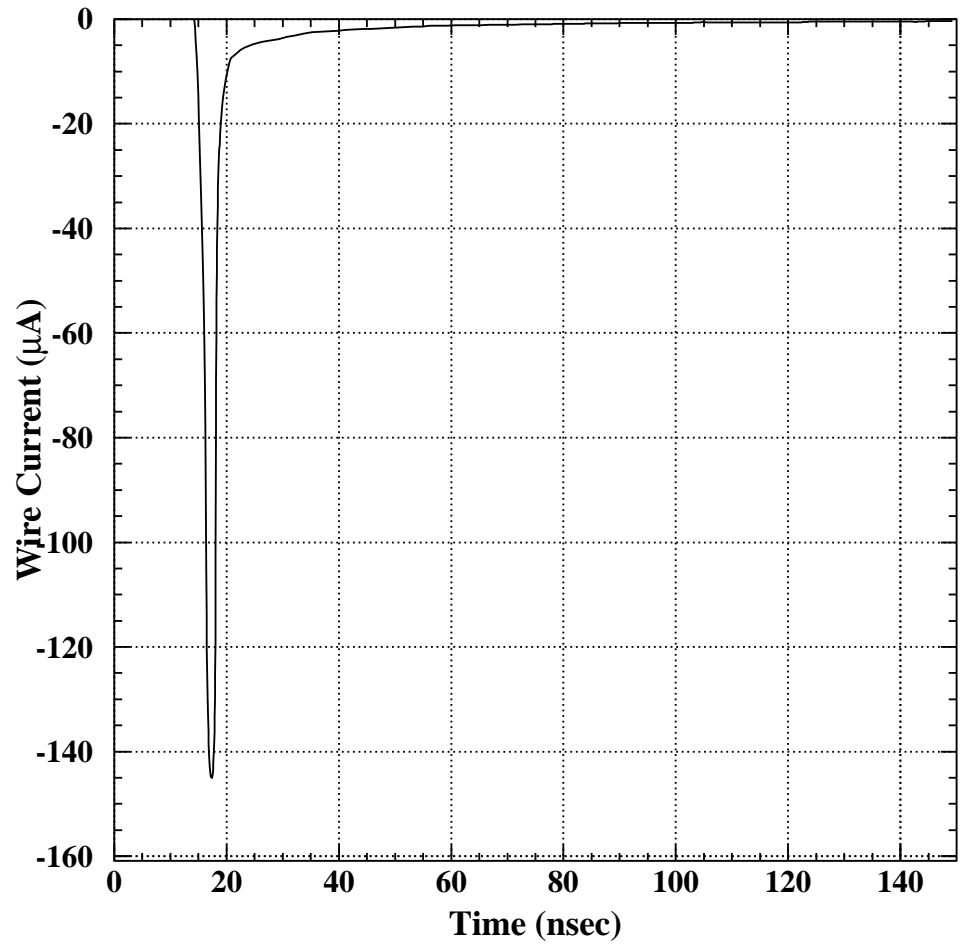}}
\caption{Anode current signal 
produced by a X-ray of $5.9~keV$ absorbed in a tube.}
\label{fecur}
\end{figure}

\begin{figure}[h]
\resizebox{14.0cm}{14.0cm}{\includegraphics{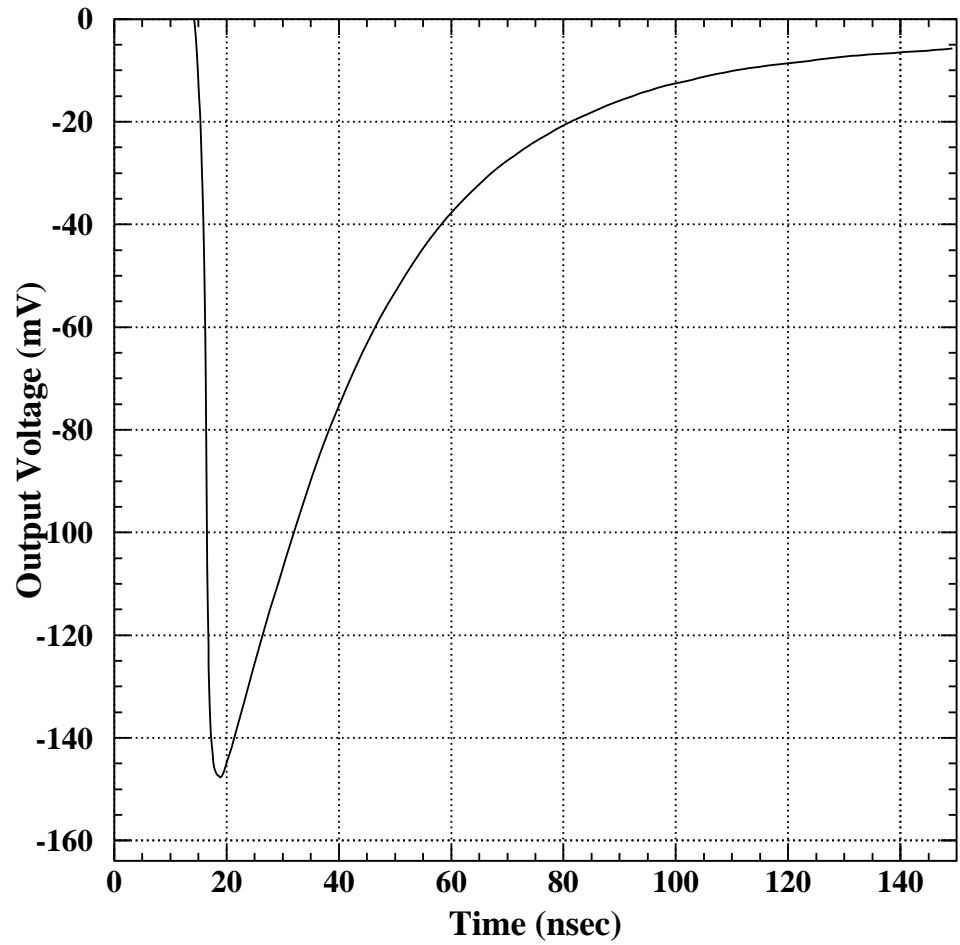}}
\caption{Output amplitude voltage
produced by a X-ray of $5.9~keV$ absorbed in a tube as processed by
the low band-pass electronic.}
\label{fevol}
\end{figure}

In Fig. \ref{fecur} a typical anode signal from a tube
produced by a X-ray of $5.9~keV$ (\nuc{55}{Fe}) is shown. When this signal is
processed by the low band-pass it assumes the shape 
reported in Fig. \ref{fevol}.
From this figure one can see that
the electronics performed a formation of the input signal with a
FWHM of about $25~nsec$. 

%\begin{figure}[h]
%\resizebox{14.0cm}{9.0cm}{\includegraphics{ioncur.eps}}
%\caption{Anode current signal
%produced by a charged particle crossing a tube.}
%\label{ioncur}
%%\end{figure}
%%\begin{figure}[h]
%\resizebox{14.0cm}{9.0cm}{\includegraphics{ionvol.eps}}
%\caption{Output amplitude voltage
%produced by a charged particle crossing a tube as processed by
%the low band-pass electronic.}
%\label{ionvol}
%\end{figure}

\begin{figure}[h]
\resizebox{14.0cm}{14.0cm}{\includegraphics{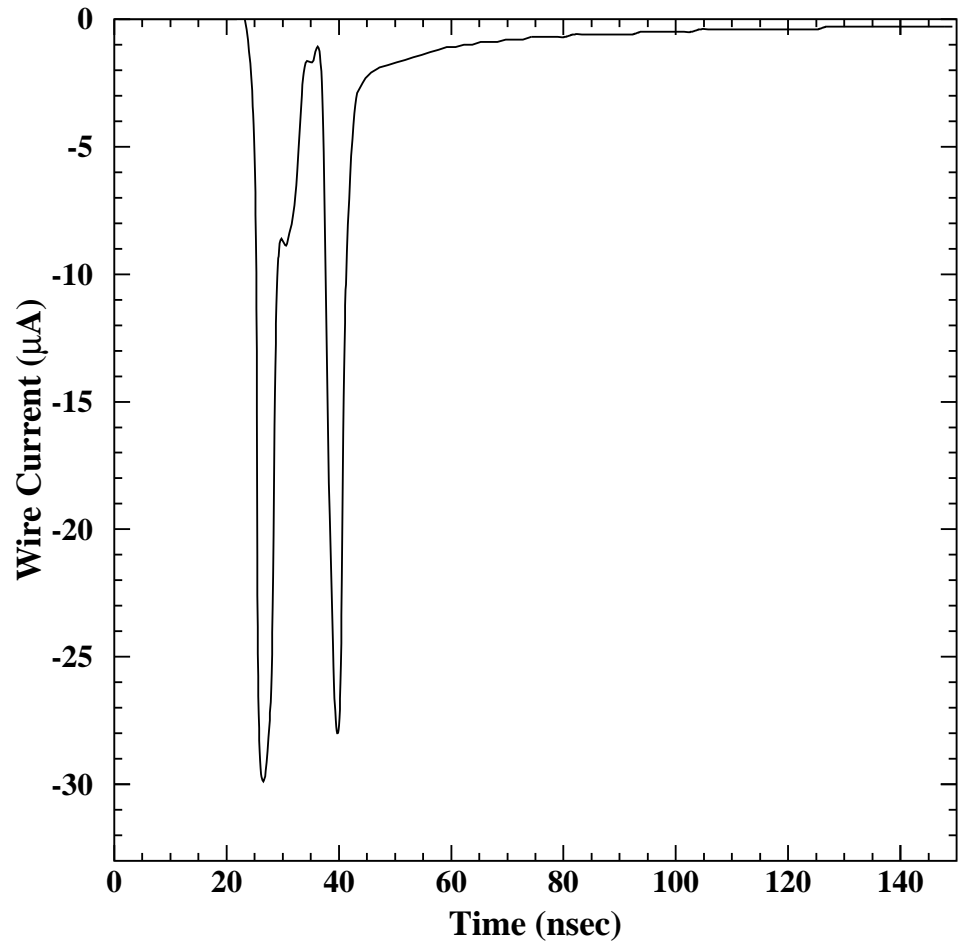}}
\caption{Anode current signal
produced by a charged particle crossing a tube.}
\label{ioncur}
\end{figure}

\begin{figure}[h]
\resizebox{14.0cm}{14.0cm}{\includegraphics{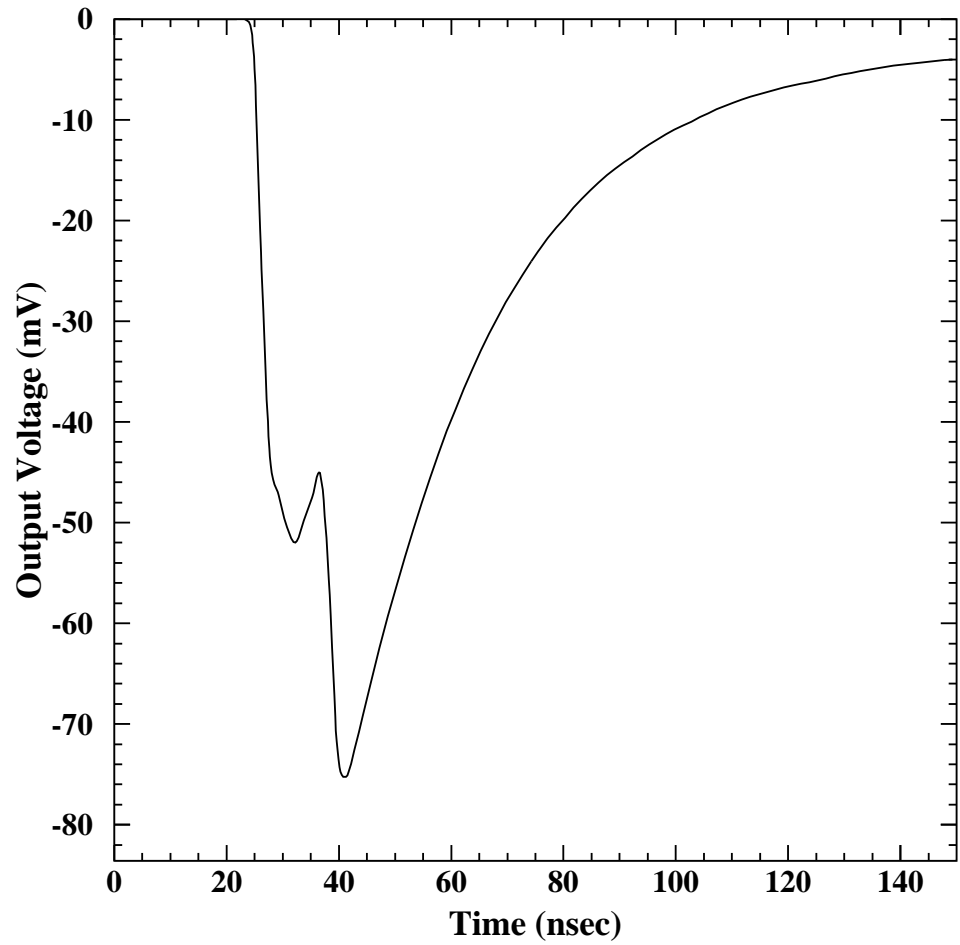}}
\caption{Output amplitude voltage
produced by a charged particle crossing a tube as processed by
the low band-pass electronic.}
\label{ionvol}
\end{figure}

In Fig. \ref{ioncur} is shown a typically anode signal produced by a
charged particle crossing a tube. In this figure one can see two peaks
are produced by two clusters. The low band-pass cannot allow to 
distinguish the two clusters since the second one is superimposed to first one
(signal pile-up) as shown in Fig. \ref{ionvol}, because their time
distance is lower than the FWHM of the electronic resolution.

\subsection{Results}
\begin{figure}[h]
\resizebox{14.0cm}{14.0cm}{\includegraphics{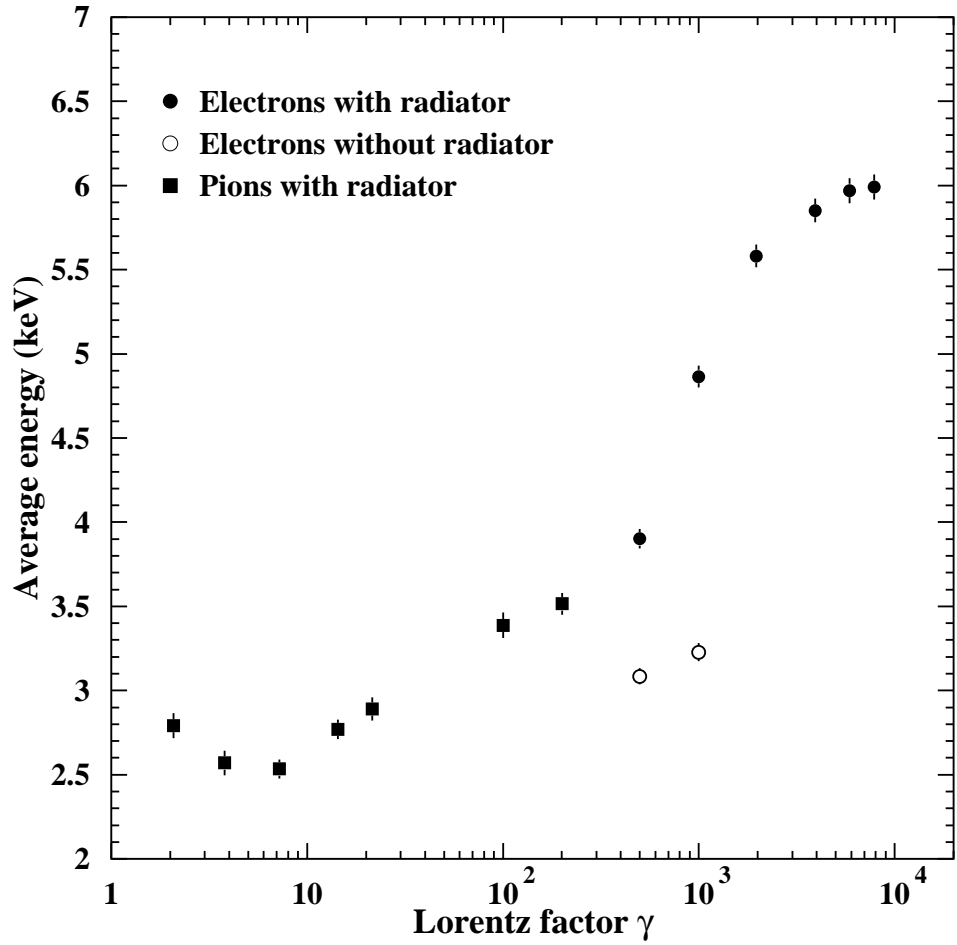}}
\caption{Average energy loss (summed over 10 planes) 
as a function of the Lorentz factor. The error bars have been
evaluated as ratio of the RMS over the square root of the number of
events.}
\label{chana}
\end{figure}

In this paragraph the results achieved by the TRD geometry defined above
are shown. In Fig. \ref{chana} the average energy loss (summed over 10
planes) as function of the
Lorentz factor is shown. This result has been obtained 
by simulating pions and
electrons of different energies with or without radiators. For each
energy 1000 events have been simulated. 

In this figure one can see that
the yield increases with $\gamma$ when the radiators
are arranged before the proportional tubes. The TR saturation is
achieved at $\gamma \simeq 8000$. For $\gamma$ less than 100--500 only the
ionization is released in the gas, as is shown in the same figure.

%\begin{figure}[h]
%\resizebox{14.0cm}{9.0cm}{\includegraphics{ener_10_5.eps}}
%\caption{Average energy loss (summed over 10 planes) 
%as a function of the Lorentz factor. The error bars have been
%evaluated as ratio of the RMS over the square root of the number of
%events.}
%\label{chana}
%\end{figure}
%
%\begin{figure}[h]
%\resizebox{14.0cm}{9.0cm}{\includegraphics{ene_el_pi.eps}}
%\caption{Average energy loss (summed over 10 planes) distribution
%for two $\gamma$ values. Solid line: pions of $255~MeV/c$; 
% dashed line: electrons of $4~GeV/c$}
%\label{peene}
%\end{figure}

\begin{figure}[h]
\resizebox{14.0cm}{14.0cm}{\includegraphics{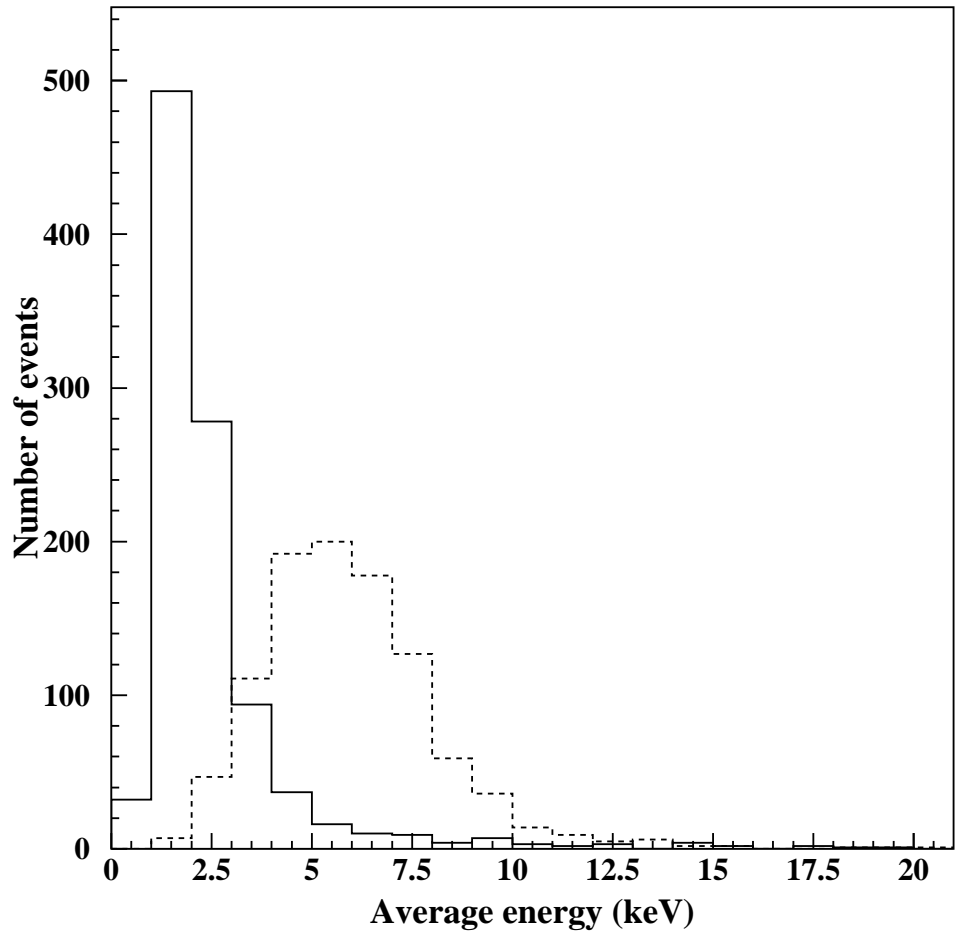}}
\caption{Average energy loss (summed over 10 planes) distribution
for two $\gamma$ values. Solid line: pions of $255~MeV/c$; 
 dashed line: electrons of $4~GeV/c$}
\label{peene}
\end{figure}

In Fig.~\ref{peene} the average energy loss distributions
(summed over 10 planes) for electrons of $4~GeV/c$ and pions of
$255~MeV/c$ are shown. From this figure it is possible to see that the
average
value of the electron distribution is greater than the average for
pions. This is due to presence of the X-ray TR produced in the
radiator by the electrons.

\begin{figure}[h]
\resizebox{14.0cm}{14.0cm}{\includegraphics{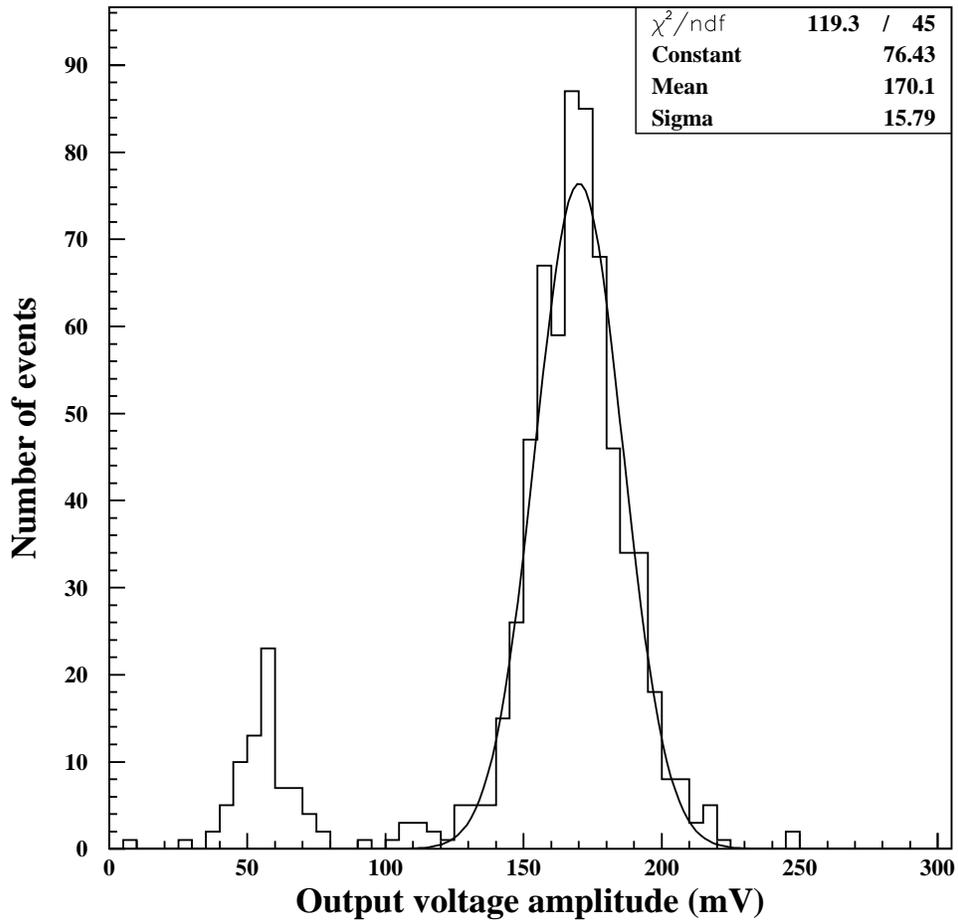}}
\caption{Output voltage amplitude distribution (histogram) 
produced by X-rays of $5.9~keV$. The line is the result of a Gaussian fit}
\label{fedist}
\end{figure}

In order to perform the cluster size analysis, one needs to know
the relationship between the output signal amplitude and the energy
loss in the tube. Therefore an analysis of voltage amplitude has
been 
done using X-rays of $5.9~keV$ (\nuc{55}{Fe}). In Fig. \ref{fedist} is
shown the output voltage amplitude distribution produced by a
\nuc{55}{Fe} X-rays absorbed in a proportional tube. 
From this figure one can see that the energy loss of $5.9~keV$
corresponds to $170~mV$ of output voltage amplitude.  

%\begin{figure}[h]
%\resizebox{14.0cm}{9.0cm}{\includegraphics{fe55_dist.eps}}
%\caption{Output voltage amplitude distribution (histogram) 
%produced by X-rays of $5.9~keV$. The line is the result of a Gaussian fit}
%\label{fedist}
%\end{figure}

\begin{figure}[h]
\resizebox{14.0cm}{14.0cm}{\includegraphics{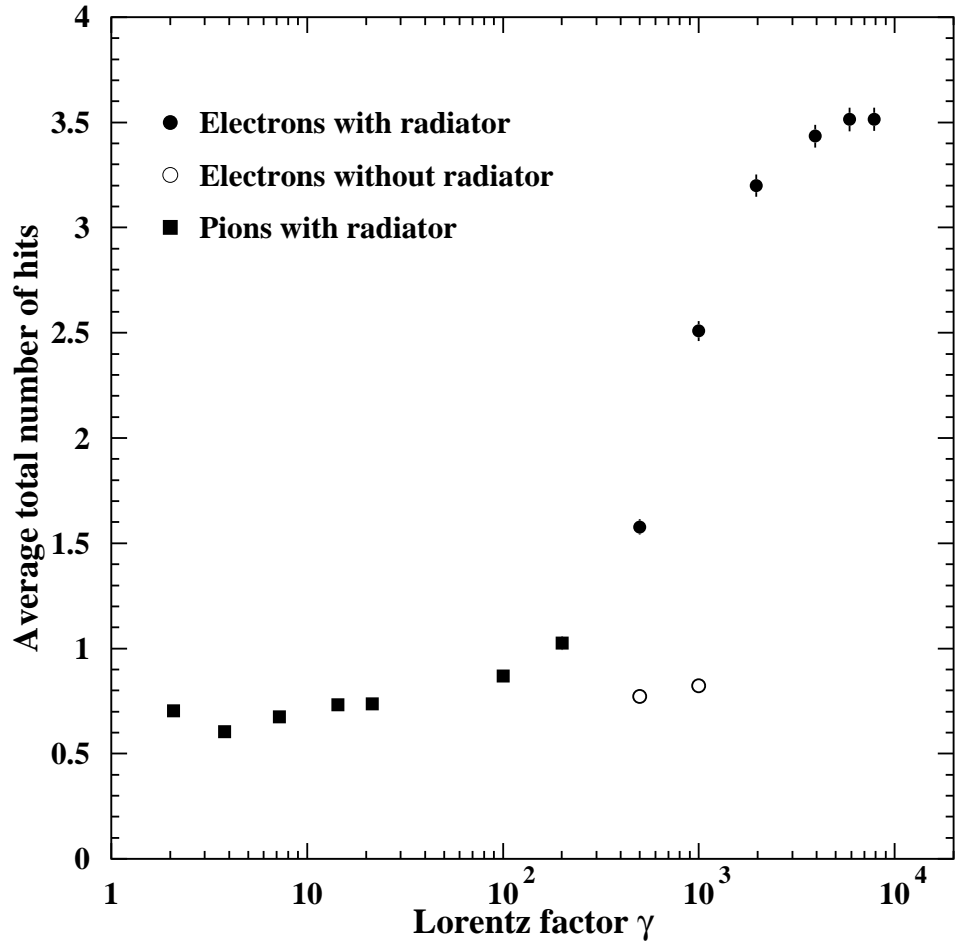}}
\caption{Total number of hits with a signal greater than $145~mV$
as a function of the Lorentz factor.
The error bars have been
evaluated as ratio of the RMS over the square root of the number of
events.}
\label{hiana}
\end{figure}

In order to count the number of hits produced for instance by TR
photons and by $\delta$-ray with energy greater than $5~keV$, a cut of 
$145~mV$ is imposed to the voltage amplitude
signal produced in each tube. In Fig. \ref{hiana} the average total number of
hits (summed over all fired tubes) when the output signal is greater 
than $145~mV$ as function of $\gamma$ is shown. The behaviour of
the TRD when is analyzed by the cluster counting method is similar to the
charge measurement one. 

\begin{figure}[h]
\resizebox{14.0cm}{14.0cm}{\includegraphics{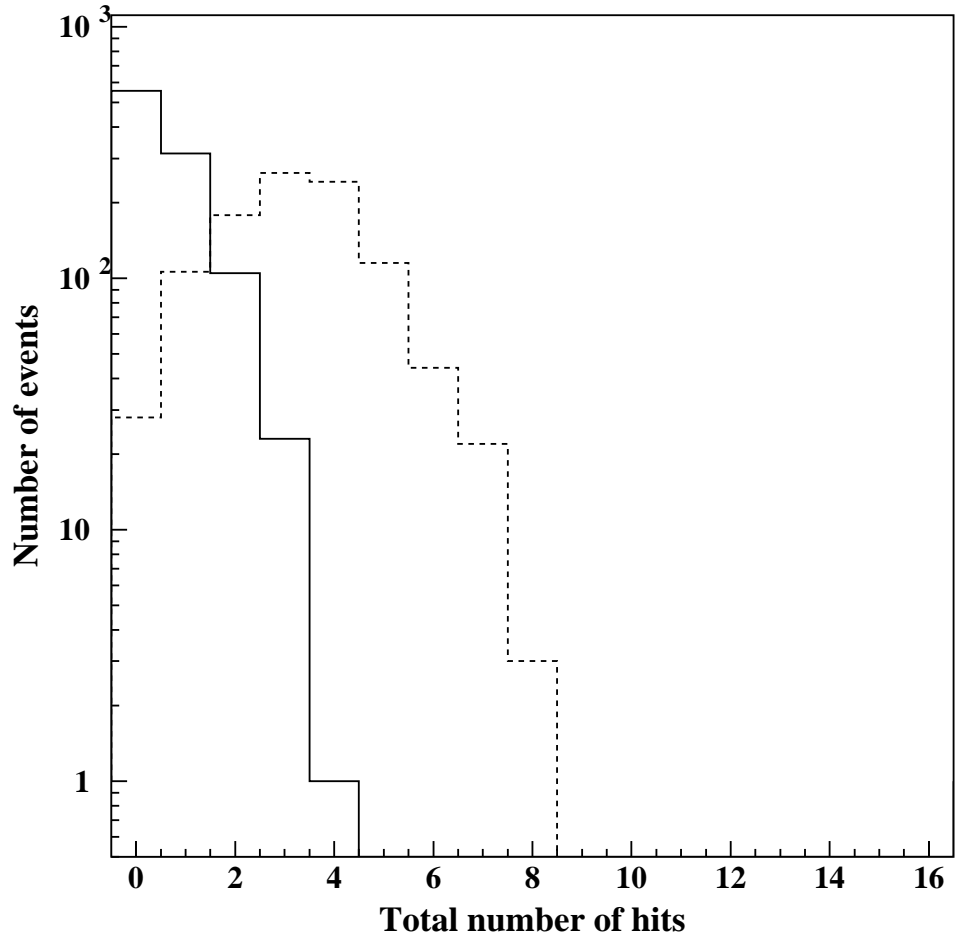}}
\caption{Hits distribution
for two $\gamma$ values. Solid line: pions of $255~MeV/c$; 
 dashed line: electrons of $4~GeV/c$}
\label{pehit}
\end{figure}

In Fig.~\ref{pehit} the distributions of the total number of hits
for electrons of $4~GeV/c$ and pions of $255~MeV/c$ are shown. 
Again  we can observe that the average value of
the electron distribution is greater than the one of the
pion distribution, due to presence of the X-ray TR produced in the
radiator by the electrons.

%\begin{figure}[h]
%\resizebox{14.0cm}{9.0cm}{\includegraphics{hits_10_5.eps}}
%\caption{Total number of hits with a signal greater than $145~mV$
%as a function of the Lorentz factor.
%The error bars have been
%evaluated as ratio of the RMS over the square root of the number of
%events.}
%\label{hiana}
%\end{figure}
%
%\begin{figure}[h]
%\resizebox{14.0cm}{9.0cm}{\includegraphics{hit_el_pi.eps}}
%\caption{Hits distribution
%for two $\gamma$ values. Solid line: pions of $255~MeV/c$; 
% dashed line: electrons of $4~GeV/c$}
%\label{pehit}
%\end{figure}

In order to discriminate electrons from pions at given momentum by
charge measurement or by cluster counting, we can
use this simulation to optimize the gas thickness, the radiator, the
threshold and the number of modules. In this way, we can optimize one
of these methods or we can use more sophisticated ones, for example
analyzing the pulse shape as function of the drift time or using the
likelihood and/or neural network analysis by the pattern information,
namely the fired tube configuration in the TRD.

\section{Conclusions}
A full simulation of a transition radiation detector (TRD) based
on the GEANT, GARFIELD, MAGBOLTZ and HEED codes has been developed. 
The simulation can be
used to study and develop TRD for high energy particle identification
using either the cluster counting or the total charge measurement method. 
The program works very well according to the design expectations.
It is quite flexible and it can be used to simulate any
detector which is based on proportional counters, providing a very
useful simulation tool.

\ack
I am grateful to Prof. P. Spinelli for useful discussions, suggestions 
and continuous support. I would like to thank my colleagues of Bari 
University and INFN for their contributions.

\newpage
\listoffigures

\newpage

\newpage

\newpage

\newpage

\newpage

\newpage

\newpage

\newpage

\newpage

\newpage

\newpage

\newpage

\end{document}